\documentclass[12pt,preprint]{aastex}
\usepackage{apjfonts}
\usepackage{epsfig}
\usepackage{amsmath}
\begin{document}

\title{Constraints on the Bulk Lorentz Factors of GRB X-Ray Flares}
\author{Shuang-Xi Yi$^{1,2}$, Xue-Feng Wu$^{3,4,5}$, Fa-Yin Wang$^{1,2}$ and Zi-Gao Dai$^{1,2}$}
\affil{$^{1}$School of Astronomy and Space Science, Nanjing University, Nanjing 210093, China; dzg@nju.edu.cn \\
$^{2}$Key Laboratory of Modern Astronomy and Astrophysics (Nanjing
University), Ministry of Education, China \\
$^{3}$Purple Mountain Observatory, Chinese Academy of Sciences,
Nanjing 210008, China\\
$^{4}$Chinese Center for Antarctic Astronomy, Chinese Academy of
Sciences, Nanjing, 210008, China\\
$^{5}$Joint Center for Particle Nuclear Physics and Cosmology of
Purple Mountain Observatory-Nanjing University, Chinese Academy of
Sciences, Nanjing 210008, China}

\begin{abstract}
X-ray flares were discovered in the afterglow phase of gamma-ray bursts
(GRBs) by the {\em Swift} satellite a decade ago and known
as a canonical component in GRB X-ray afterglows.
In this paper, we constrain the Lorentz factors of
GRB X-ray flares using two different methods. For the first
method, we estimate the lower limit on the bulk Lorentz factor with
the flare duration and jet break time. In the second method, the
upper limit on the Lorentz factor is derived by assuming that the
X-ray flare jet has undergone saturated acceleration. We also
re-estimate the initial Lorentz factor with GRB afterglow onsets, and find the
coefficient of the theoretical Lorentz factor is 1.67 rather than the commonly used 2 for interstellar
medium (ISM) and 1.44 for the wind case. We find that the correlation
between the limited Lorentz factor and the isotropic radiation
energy of X-ray flares in the ISM case is more consistent with that
of prompt emission than the wind case in a statistical sense.
For a comparison, the lower limit on Lorentz factor is statistically larger
than the extrapolation from prompt bursts in the wind case. Our
results indicate that X-ray flares and prompt bursts are produced by
the same physical mechanism.
\end{abstract}


\section{Introduction}

Gamma-ray bursts (GRBs) are the most luminous explosive events in
the Universe. It is well known that the observed gamma-ray emission
are produced by relativistic outflows. The fireball of a GRB is required to have a
relativistic speed toward the earth in order to avoid the
``compactness problem''. The Lorentz factor of the fireball increases
with radius in the radiation-dominated acceleration phase. At the
end of this phase, the fireball enters the matter-dominated coasting
phase, which maintains a constant Lorentz factor (for reviews see Piran 1999;
Zhang \& M\'{e}sz\'{a}ros 2004; M\'{e}sz\'{a}ros 2006; Kumar \& Zhang 2015).
The Lorentz factor in the matter-dominated coasting phase, called the initial
Lorentz factor, is a crucial parameter in understanding the physics
of GRBs. There are several methods to constrain this initial Lorentz
factor $\Gamma_0$, but the most effective method is that the
peak time of a GRB early afterglow onset is taken as the deceleration
time of the outflow. An estimation of this initial value is possible by
measuring the peak time of the afterglow light curve and the prompt isotropic
energy $E_{\gamma ,iso}$ (e.g., Molinari et
al. 2007). This method was successfully applied in Liang et al.
(2010) with a large sample of the early afterglow light curves. Liang et al. (2010)
discovered a tight correlation between $\Gamma_0$ and $E_{\gamma
,iso}$. In this paper, we re-estimate the initial Lorentz factor of
GRBs in the ISM and wind cases, and also obtain a tight
correlation between $\Gamma_0$ and $E_{\gamma ,iso}$. The $\Gamma_0$
and $E_{\gamma ,iso}$ correlation in the wind case is even more
tighter than that of the ISM case.

X-ray flares are common phenomena in GRB X-ray afterglows in the {\em
Swift} era. According to the temporal behavior and spectral properties of X-ray flares, it is widely argued that X-ray
flares are produced by long-lasting central engine activities (Burrows
et al. 2005; Ioka et al. 2005; Fan \& Wei 2005; Falcone et al. 2006;
Zhang et al. 2006; Wang \& Dai 2013). It is generally believed that
X-ray flares have the same physical origin as the prompt emission of
GRBs. X-ray flares usually happen at $\sim 10^2-10^5$ s after prompt
emissions, transferring a large percentage of the outflow energy to
the radiation. Most of the methods proposed to constrain the
initial Lorentz factor of GRBs are not suitable for X-ray flares.
Since the fluence of most X-ray flares are smaller than the prompt emission,
their energies and Lorentz factors are supposed to be smaller than those of GRBs.
In this paper, we use two different methods to place limits on the
Lorenz factor of X-ray flares. In Section 2, we introduce the method
to estimate of the initial Lorentz factor $\Gamma_0$ of GRBs. The
two methods of constraining the X-ray flares Lorentz factors are
shown in Section 3. Sample study are presented in Section 4, and our
results are summarized and discussed in Section 5.  A concordance
cosmology with parameters $H_0 = 71$ km s$^{-1}$ Mpc$^{-1}$,
$\Omega_M=0.30$, and $\Omega_{\Lambda}=0.70$ are adopted. Notation
$Q_n$ denotes $Q/10^n$ in the cgs units throughout the paper.

\section{Estimating the Initial Lorentz Factor $\Gamma_0$}
When a relativistic fireball shell sweeps up the circumburst medium,
two shocks appear: a reverse shock propagating into the fireball
shell, and a forward shock propagating into the ambient medium. We
suppose that the fireball shell and two shocks are spherical.
Physical quantities denoted by
``${\prime}$'' are defined in the co-moving frame. We divide this
two-shock system into 4 regions (Sari \& Piran 1995; Yi et al.
2013): (1) The unshocked ambient medium ($n_1$, $e_1$, $p_1$,
$\Gamma_1$), (2) the shocked ambient medium ($n'_2$, $e'_2$, $p'_2$,
$\Gamma_2$), (3) the shocked fireball shell ($n'_3$, $e'_3$, $p'_3$,
$\Gamma_3$), and (4) the unshocked fireball shell ($n'_4$,
$\Gamma_4=\Gamma_0$), where $n$ is the number density, $e$ is the
internal energy density, $p$ is the pressure, and $\Gamma$ is the
bulk Lorentz factor. Hereafter we calculate the properties of the
reverse shock emission at the radius $R_{\times}$, where the reverse shock
finishes crossing the ejecta shell and the Lorentz factor of the shell is $\Gamma_\times$.

The assumptions of the equilibrium of pressures and equality of velocities along the
contact discontinuity lead to $p'_2=p'_3$ and $\Gamma_2=\Gamma_3$,
respectively. With the jump condition for the shocks and the
equilibrium of pressures, we can obtain,
\begin{equation}
(4{\Gamma _{34}} + 3)({\Gamma _{34}} - 1){n'_4} = (4{\Gamma _2} +
3)({\Gamma _2} - 1){n_1} .
\end{equation}
The Lorentz factor of the reverse shock $\Gamma_{34}$ could be approximated as
\begin{equation}
{\Gamma _{34}} = \frac{1}{2}\left( {\frac{{{\Gamma _3}}}{{{\Gamma _4}}} + \frac{{{\Gamma _4}}}{{{\Gamma _3}}}} \right) = \frac{1}{2}\left( {\frac{{\Gamma _4^2 + \Gamma _3^2}}{{{\Gamma _4}{\Gamma _3}}}} \right) ,
\end{equation}
as long as $\Gamma_{3} \gg 1$ and $\Gamma_{4} \gg 1$.
Substituting Equation (2) into Equation (1), we can obtain the following equation
\begin{equation}
\frac{{{{({\Gamma _4} - {\Gamma _3})}^2}}}{{{\Gamma _4}{\Gamma _3}}}\left[ {\frac{{{{({\Gamma _4} + {\Gamma _3})}^2}}}{{{\Gamma _4}{\Gamma _3}}} - \frac{1}{2}} \right]{n'_4} = 4\Gamma _3^2{n_1} .
\end{equation}
Because $\Gamma_{3} \gg 1$ , $\Gamma_{4} \gg 1$ and $\Gamma_{4} \geq
\Gamma_{3}$, we ignore the constant 1/2 term in the Equation (3) and thus
we can obtain the solution of this equation (ignoring the
negative solution, also see Panaitescu \& Kumar 2004)
\begin{equation}
{\Gamma _3} = \frac{{{\Gamma _4}}}{{{{\left[ {1 + 2{\Gamma _4}{{\left( {{n_1}/{n'_4}} \right)}^{{1 \mathord{\left/
 {\vphantom {1 2}} \right.
 \kern-\nulldelimiterspace} 2}}}} \right]}^{{1 \mathord{\left/
 {\vphantom {1 2}} \right.
 \kern-\nulldelimiterspace} 2}}}}} .
 \end{equation}
Here we obtain the relation between the Lorentz factor of shocked
fireball shell $\Gamma_3$ and the initial Lorentz factor $\Gamma_0$ ($\Gamma_4$),
which depends on the ratio of these two comoving densities. The
number density of the ambient medium is assumed to be $n_1 =
AR^{-k}$ (Dai \& Lu 1998; M\'esz\'aros et al 1998; Chevalier \& Li
2000; Wu et al 2003, 2005;  Yi et al. 2013), such a
circumburst medium is a homogeneous interstellar medium (ISM) for
$k=0$, and a typical stellar wind environment for $k=2$. The fireball
shell is characterized by an initial kinetic energy $E_k$, initial
Lorentz factor $\Gamma_{4}$, and a width $\Delta$ in the lab frame
attached to the explosion center, so the number density of the shell
in the comoving frame is ${n'_4} = E_{k}/(4\pi
{m_p}{c^2}{R^2}\Delta{\Gamma_{4} ^2})$. The ratio of the comoving number
density of the relativistic shell ${n'_4}$ to the number density of
the ambient medium ${n_1}$ defined in Sari \& Piran (1995) is
\begin{equation}
f = \frac{{n'_4}}{{{n_1}}} = \frac{E_k}{{4\pi A{m_p}{c^2}\Delta \Gamma _4^2{R^{2 - k}}}} = \frac{X}{{\Delta \Gamma _4^2{R^{2 - k}}}},
\end{equation}
where $X = E_{k}/(4\pi A {m_p}{c^2})$.
The difference between the lab frame speed of the unshocked fireball shell and that of the reverse shock is (Kumar \& Panaitescu 2003),
\begin{equation}
{\beta _4} - {\beta _{RS}} = \frac{{1.4}}{{\Gamma _4^2}}{\left( {\frac{{\Gamma _4^2{n_1}}}{{n'_4}}} \right)^{\frac{1}{2}}} = \frac{{1.4}}{{{\Gamma _4}}}{\left( {\frac{1}{f}} \right)^{\frac{1}{2}}} .
\end{equation}
Considering the thin shell case $\bigtriangleup \simeq R/(2{\Gamma
_4^2})$, we can calculate the radius $R_{\times}$ where the reverse
shock finishes crossing the fireball shell,
\begin{equation}
\Delta ({R_ {\times}}) = \int_0^{{R_ {\times} }} {\left( {{\beta _4} - {\beta _{RS}}} \right)} dR .
\end{equation}
The substitution of Equation (5) and (6) into Equation (7) leads to
\begin{equation}
{R_ \times } = {\left[ {\frac{{2{{(5 - k)}^2}X}}
{{{{5.6}^2}\Gamma _4^2}}} \right]^{\frac{1}
{{3 - k}}}} = {\left[ {\frac{{{{(5 - k)}^2}{E_k}}}
{{2 \times {{5.6}^2}\pi A{m_p}{c^2}\Gamma _4^2}}} \right]^{\frac{1}
{{3 - k}}}}.
\end{equation}
So the comoving density ratio at $R_\times$ is
\begin{equation}
{f_ \times } = \frac{{n_4^{'}}}
{{{n_1}}} = \frac{{2X}}
{{R_ {\times} ^{3 - k}}} = \frac{{{{5.6}^2}}}
{{{{(5 - k)}^2}}}\Gamma _4^2.
\end{equation}
Substituting Equation (9) into Equation (4), we can obtain the Lorentz factor of the reverse shock as it finishes crossing the shell
\begin{equation}
{\Gamma _ \times } = \frac{{{\Gamma _4}}}{{{{\left[ {1 + 2{\Gamma _4}{{\left( R^{3 - k}/(2X) \right)}^{{1 \mathord{\left/
 {\vphantom {1 2}} \right.
 \kern-\nulldelimiterspace} 2}}}} \right]}^{{1 \mathord{\left/
 {\vphantom {1 2}} \right.
 \kern-\nulldelimiterspace} 2}}}}} = \frac{{{\Gamma _4}}}{{{{\left[ {1 + 0.357(5 - k)} \right]}^{{1 \mathord{\left/
 {\vphantom {1 2}} \right.
 \kern-\nulldelimiterspace} 2}}}}} .
\end{equation}
Therefore, the relation between $\Gamma _\times$ and the initial Lorentz factor $\Gamma _0$ is
\begin{equation}
{\Gamma _ \times }=0.60\,{\Gamma _ 0 },\;\; {\rm for} \;\;k = 0 \;({\rm ISM}),
\end{equation}
and
\begin{equation}
{\Gamma _ \times }=0.70\,{\Gamma _ 0 },\;\;{\rm for} \;\;k = 2 \;({\rm Wind}).
\end{equation}
For the thin shell case, the reverse shock crossing time $T_\times$ is
almost corresponding to the deceleration time $T_{dec}$, .i.e,
$T_{\times} \sim T_{dec}$. Therefore, we can derive the initial Lorentz
factor in the ISM and wind type cases (also see Panaitescu \& Kumar
2004). For $k=0$ (ISM),
\begin{equation}
{\Gamma _0} = \frac{1}{{0.60}}{\Gamma _ \times } = 1.67{\left[ {\frac{{3{E_{\gamma ,iso}}}}{{32\pi {n_1}{m_p}{c^5}\eta t_{p,z}^3}}} \right]^{\frac{1}{8}}},
\end{equation}
and for $k=2$ (wind),
\begin{equation}
{\Gamma _0} = \frac{1}{{0.70}}{\Gamma _ \times } = 1.44{\left[ {\frac{{{E_{\gamma ,iso}}}}{{8\pi A{m_p}{c^3}\eta {t_{p,z}}}}} \right]^{\frac{1}{4}}}.
\end{equation}
With the isotropic-equivalent energy $E_{\gamma ,iso}$ and the peak time of the afterglow onset $t_{p,z}$,
we can estimate the initial Lorentz factor of GRBs, where $t_{p,z} =
t_{p}/(1+z)$. Liang et al. (2010) discovered a tight correlation
between $\Gamma_0$ and $E_{\gamma ,iso}$ using 20 GRBs which show
deceleration feature in the early afterglow light curves. Other work
also confirmed this correlation, but with different methods and
power-law indices (Ghirlanda et al. 2012; L{\"u} et al. 2012). Using the
data of $t_{p,z}$ and $E_{\gamma ,iso}$ from Liang et al. (2010, 2013) and L{\"u} et al. (2012), we
re-constrain the initial Lorentz factor, and also discover a tight
$\Gamma_0$ and $E_{\gamma ,iso}$ correlation for the ISM and wind cases.
The $\Gamma_0$ and $E_{\gamma ,iso}$ correlation in the wind case is
even more tighter than that in the ISM case, as shown in Figs. 4 and 5.

\section{Methods of Constraining the Bulk Lorentz Factor of X-ray flares}
Because most of the models estimating the bulk Lorentz factor during the GRB
prompt emission phase are inapplicable for the X-ray flares, we introduce two
methods to constrain the upper and lower limits on the Lorentz factor of X-ray flares in
this section. In principle, GRB outflows could be structured, but we suppose that the outflows are conical (also called jet-like) and
the half-opening angles of the outflows have a same/constant value in a single GRB, i.e., the half-opening angle of the X-ray flare jet is the same as that of the prompt jet in one GRB.
Therefore, we suppose that each X-ray flare is produced in a
conical uniform jet with half-opening angle $\theta_{j}$. If there
are several flares in one X-ray afterglow, we also assume these
X-ray flares have the same jet half-opening angle $\theta_{j}$.

{\bf Method I}: Lower limit on the Lorentz factor of X-ray flares.
We use the late internal shock emission model to constrain the lower
limits of the Lorentz factor of X-ray flares (Wu et al. 2006, 2007).
The quick decline of X-ray flares after the peak time is widely
interpreted as the high latitude component of X-ray pulses (Burrows et
al. 2005; Tagliaferri et al. 2005; Zhang et al. 2006; Nousek et al.
2006; Liang et al. 2006). Wu et al. (2007) supposed that emission
from the same internal shock radius $R_{int}$ but with different
angles would have different arrival times to the observer, which is due
to light propagation effect. The delay time of different photons
with different $\theta$ $(\theta < \theta_{j})$ is
\begin{equation}
\Delta T = \frac{{{R_{{\mathop{\rm int}} }}(1 - \cos \theta )}}{c}.
\end{equation}
If photons are emitted at an angle $\theta = \theta_{j}$, then the
delay time $\Delta T$ is about the duration time of the
X-ray flare. We denote the timescale of the decay part of
an X-ray flare by $T_{decay}$. We could put forth
a constraint on the decaying timescale of the flare and the
jet-opening angle, i.e.
\begin{equation}
{T_{decay}} < \frac{{{R_{{\mathop{\rm int}} }}(1 - \cos \theta_{j} )}}{c}.
\end{equation}
Another timescale is the angular spreading variability time $T_{rise}$, which is the rise time of an X-ray flare,
\begin{equation}
{T_{rise}} = \frac{{{R_{{\mathop{\rm int}} }}(1 - \cos (1/{\Gamma _x}))}}{c} \approx \frac{{{R_{{\mathop{\rm int}} }}}}{{2\Gamma _x^2c}}.
\end{equation}
Combining the decaying timescale and rising timescale, we could get a
lower limit on the Lorentz factor of each X-ray flare,
\begin{equation}
{\Gamma _x} > {\left( {\frac{{{T_{decay}}}}{{{T_{rise}}}}} \right)^{\frac{1}{2}}}{\left [ {\frac{1}{{2(1 - \cos {\theta _{j}})}}} \right]^{\frac{1}{2}}} \approx \theta _{j}^{ - 1}{\left( {\frac{{{T_{decay}}}}{{{T_{rise}}}}} \right)^{\frac{1}{2}}}.
\end{equation}
With the jet opening angle $\theta_{j}$, decaying timescale and
rising timescale, we can constrain the lower limit of the Lorentz
factor of each X-ray flare. The jet half-opening angle $\theta_{j}$ can
be estimated from the late afterglow of a GRB if there is a jet
break in the afterglow light curve. So, we selected GRBs which have
jet breaks and flares in the X-ray afterglow light curve in our
sample. The sample is listed in Table 1.

{\bf Method II}: Upper limit on the Lorentz factor of X-ray flares.
The physical mechanism of X-ray flares is still unclear. X-ray
flares are known to be similar to those of prompt emission pulses
through studying of the temporal behavior and energy spectrum
(Chincarini et al. 2010). We generally consider X-ray flares having
the same physical origin as the prompt emission of GRB, and they are
all due to the long-lasting activity of the central engine (Fan \& Wei 2005;
Zhang et al. 2006). But there is still controversy about the origin of X-ray flares of
GRBs. X-ray flares of short GRBs can be produced by
differentially rotating, millisecond pulsars from
the mergers of binary neutron stars (Dai et al. 2006). Magnetic
reconnection-driven explosions lead to multiple X-ray flares minutes
after the prompt GRB. Wang \& Dai (2013) performed a statistical
study of X-ray flares for long and short GRBs, and found
energy, duration, and wait-time distributions similar to those of solar flares, which
indicates that X-ray flares of GRBs may be powered by magnetic
reconnection. According to the standard GRB fireball model, after the initial
radiation-dominated acceleration phase, the fireball enters the
matter-dominated ``coasting'' phase (Piran 1999; M\'{e}sz\'{a}ros
2002; Zhang \& M\'{e}sz\'{a}ros 2004). Whether the fireball is baryon-rich or not
determines how long is the initial radiation-dominated
acceleration phase. If the fireball is baryon-poor, the initial
energy of the fireball will be quickly converted into radiation energy and
produce bright and brief thermal emission, which is inconsistent with most
of the observations. The spectra of X-ray flares are typically non-thermal,
with a photon index of about $\sim-2.0$ (Falcone et al. 2007). This
suggests that X-ray flares happen when the jet is optically thin.
Meanwhile, thermal emission just before any X-ray flare with
compatible flux has never been detected, indicating that the jet
responsible for the flare attains saturated acceleration. This
requires baryon loading in the jet to be large enough, or that the
Lorentz factor has an upper limit. On the other hand, the upper
limit on the Lorentz factor of an X-ray flare can be estimated as
(Jin et al. 2010)
\begin{equation}
{\Gamma _x} \le {\left( {\frac{{L\,{\sigma _T}}}{{8\pi {m_p}{c^3}{R_0}}}} \right)^{\frac{1}{4}}},
\end{equation}
which depends on the total luminosity $L$ and initial radius $R_0$
of the flare outflow. Jin et al. (2010) assumed that the observed
X-ray flare luminosity is just a fraction ($\epsilon_x=0.1$) of the
total luminosity of the outflow, that is $L_{x} = 0.1 L$ (also see,
Fan \& Piran 2006). $R_{0}$ is taken to be $10^{7}$ cm, which is
comparable to the radius of a neutron star. $R_{0} = 10^{7}$ cm is a
conservative value, in some cases, $R_{0}$ is taken to be $\sim
10^{8}$ cm or even larger (Pe'er et al. 2007). With the proper
values, we are able to get an upper limit on $\Gamma_{x}$. In our
sample, $L_{x}$ is taken as $ E_{x,iso}/T_{90,x}$, where $E_{x,iso}$
and $T_{90,x}$ are the isotropic energy and duration time of one
flare respectively. The redshift of those GRBs in our sample are
all measured, the isotropic 0.3 - 10 keV energy of the X-ray flares in
the sample can be estimated from the fluence as
\begin{equation}
{E_{x,iso}} = \frac{{4\pi D_L^2}}{{1 + z}}\,S_x,
\end{equation}
where $S_x$ is the fluence of an X-ray flare.

\section{Case Studies}
Our method for placing lower limits on the Lorentz factor is feasible if
the X-ray light curve presents flare and jet break simultaneously.
Our sample consists of 20 GRBs with X-ray flares, redshift and jet
break time (Falcone et al. 2007; Chincarini et al. 2010; Bernardini
et al. 2011; Lu et al. 2012). Some of them have several flares in one GRB. The
total number of X-ray flares is 43. We assume that the opening angle is
the same for the jets responsible for prompt emission and late X-ray
flares in a single GRB. According to the appearance time of X-ray flares, we
mark the corresponding numerical order, which can be seen in the
Table 1. The $T_{rise}$, $T_{decay}$, $S_x$, and $T_{90,x}$ of each
flare are reported in Falcone et al. (2007), Chincarini et al.
(2010) and Bernardini et al. (2011).
$\theta_{j}^{Wind}$ and $\theta_{j}^{ISM}$ are calculated with the data taken from Lu et al.
(2012) when the Eq. (22) and Eq. (23) are applied. The Lorentz factor in the
wind type circumburst media is (Chevalier \& Li 2000)
\begin{equation}
\Gamma  = 5.9{\left( {\frac{{1 + z}}{2}} \right)^{{1 \mathord{\left/
 {\vphantom {1 4}} \right.
 \kern-\nulldelimiterspace} 4}}}E_{k,52}^{{1 \mathord{\left/
 {\vphantom {1 4}} \right.
 \kern-\nulldelimiterspace} 4}}\;A_*^{{{ - 1} \mathord{\left/
 {\vphantom {{ - 1} 4}} \right.
 \kern-\nulldelimiterspace} 4}}\,t_{days}^{ - {1 \mathord{\left/
 {\vphantom {1 4}} \right.
 \kern-\nulldelimiterspace} 4}} ,
\end{equation}
where $E_{k,52}$ is the initial kinetic energy of the fireball shell in
units of $10^{52}$ ergs, $t_{days}$ is the observer's time in units
of days, $ A = \mathop {{M_w}}\limits^. /4\pi {V_w} = 5 \times
{10^{11}}{A_*}\; \rm g\;c{m^{ - 1}}$ is the wind parameter, $\mathop {{M_w}}\limits^.$ is
the wind mass-loss rate, and $V_w$ is the wind velocity. Because the jet
break effects are considered when $\Gamma \approx \theta_j^{-1}$, so
Equation (21) could be used to estimate the jet half-opening angle for the wind case,
\begin{equation}
\theta_{\rm j}^{wind}=0.12\,\,{\rm rad}\,\,\left(\frac{T_{j}}{1\ \rm
day}\right)^{1/4}\left(\frac{1+z}{2}\right)^{-1/4}
E_{\gamma,iso,52}^{-1/4}\left(\frac{\eta}{0.2}\right)^{1/4}A_{*}^{1/4}.
\end{equation}
where $\eta$ is the efficiency of prompt GRBs
and $A_* = 1$ adopted in this paper. The jet half-opening angle in the interstellar medium case
could be described by (Sari et al. 1999; Rhoads 1999; Frail et al. 2001; Yi et al. 2015),
\begin{equation}
\theta_{\rm j}^{ISM}=0.076 \,\,{\rm rad}\,\,\left(\frac{T_{j}}{1\ \rm
day}\right)^{3/8}\left(\frac{1+z}{2}\right)^{-3/8}
E_{\gamma,iso,53}^{-1/8}\left(\frac{\eta}{0.2}\right)^{1/8}\left(\frac{n}{1\ \rm
cm^{-3}}\right)^{1/8}.
\end{equation}
The distribution of jet half-opening
angles for the ISM and wind cases is shown in Fig 1.

We assume that X-ray flares are coming from relativistic jets.
The tail of an X-ray flare is interpreted as emission from high
latitude areas of the jet. The duration of the flare is determined
by the half-opening angle of the jet through the curvature effect. With
the jet half-opening angle $\theta_j$ estimated from the jet break
time, decaying timescale and rising timescale of the flare, we can
obtain the lower limit on the Lorentz factor of the flare via
Equation (18). The upper limit on the Lorentz factor is
determined by the total luminosity and initial radius of the outflow. The
observed average luminosity is obtained from the isotropic 0.3 -
10 keV energy of the X-ray flare averaged by the duration time of the
flare. The total luminosity of the flare is assumed to be 10 times of the
observed X-ray luminosity. As already mentioned, the
initial radius of the outflow $R_0$ is taken as $10^7$ cm. The
obtained limits on the bulk Lorentz factor of X-ray flares range
from tens to hundreds, as can be seen in Fig 2 and 3. We find that in the ISM case the correlation
between the Lorentz factor and the isotropic radiation energy of
X-ray flares is almost consistent with that of  prompt emission of GRBs (Fig 4).
However, in the wind case the lower limit on the Lorentz factor is
statistically larger than the extrapolation from prompt bursts (Fig 5).

\section{Discussion}
X-ray flares are common features in GRB X-ray
afterglows, and most of them have occurred at the early period. We can
conclude that all the flares in our sample occurred before the jet break,
which can be seen from the Table 1. Here, we define
$f=\theta_{j,\,\gamma}/\theta_{j,\,x}$, where $\theta_{j,\,x}$ is
the half-opening angle of the X-ray flare jet while
$\theta_{j,\,\gamma}$ is the half-opening angle of
the jet responsible for prompt emission. If $f=1$, one suggests that the jet may be conical and the flare jet and prompt emission jet have the same half-opening angle as we discussed above.
The jet opening angle might be larger during the prompt emission
and smaller for the X-ray flares, i.e., $f>1$, which
has been predicted in some models with magnetic-dominated jets (Levinson \&
Begelman 2013; Bromberg et al. 2014). In this case, the lower limit of flare
Lorentz factor would be larger than that estimated with Eq. (18) assuming $\theta_{j,\,x}=\theta_{j,\,\gamma}$. The corresponding lower
limits on X-ray flare Lorentz factor in Figs. 4 and 5 will increase by a factor of $f$, making the X-ray flares possibly more inconsistent (especially for the wind case) with
the extrapolation of the correlation between isotropic radiation energy and Lorentz
factor of prompt emission of GRBs. On the other hand, although several observations suggest that in some GRBs the ejecta may have large scale magnetic fields and therefore the
ejecta could be magnetized, the degree of magnetization is usually estimated as $\sigma<$ a few
in the afterglow phase. So in this paper we assume
that for simplicity GRB jets have negligible magnetization, and the outflows have same half-opening angle ($f=1$) in one GRB.

In addition, the fluence of most X-ray flares are smaller than that of prompt emission, their energies and Lorentz factors are supposed
to be smaller than those of GRBs. The initial Lorentz factor of GRBs
in this paper is generally larger than a few hundreds, and it is always larger than the lower limits of X-ray flare Lorentz factor in the same GRB.
In Fig 6, we plot 5 GRBs having prompt and flare Lorentz factors,
GRBs 050820A, 060418, 060906, 070318, and 071031. The
initial Lorentz factors of these 5 GRBs are generally much larger
than the lower limits of the Lorentz factor of their X-ray flares, and usually smaller than
the upper limits of flare Lorentz factors.

\section{Conclusion}

The initial Lorentz factor is a key parameter to understanding the
GRB physics. In this paper, we have re-estimated the initial Lorentz
factor in a more accurate way. From Equation (13), we obtain a coefficient 1.67 for the ISM case,
instead of 2 adopted in previous literature. We also
constrain the initial Lorentz factor in the wind case, which
is shown as Equation (14). With the estimated initial Lorentz factors
in this paper, we confirm the tight correlation between the initial
Lorentz factor and isotropic energy of GRBs for the ISM case. There is an
even tighter correlation between the initial Lorentz factor and
isotropic energy of GRBs for the wind case.

Our sample consists of 20 GRBs with X-ray flares, whose redshifts and
jet break times are known. Some of them have several flares. The
total number of X-ray flares in our sample is 43. We assume that the half-opening angle is
the same for the jets responsible for prompt emission and late X-ray
flares in one GRB. Our results are shown in Fig 4 (ISM) and Fig 5
(Wind), which also show the correlation between isotropic radiation
energy and the Lorentz factor of prompt emission of GRBs. The obtained
limits on the bulk Lorentz factor of X-ray flares range from a few tens
to hundreds, together with the isotropic radiation energy, are generally
consistent with the correlation for prompt GRBs for the ISM case. Our
results indicate that X-ray flares and prompt bursts may be caused by
the same mechanism, as both are produced by the long-lasting activity of the central
engine. However, in the wind case the lower limit on Lorentz factor
is statistically larger than the extrapolation from prompt bursts.

\acknowledgements
We thank the anonymous referee for constructive suggestions.
We also thank Bing Zhang, En-Wei Liang, Yun-Wei Yu, Liang-Duan Liu, Di Xiao and A-Ming Chen for useful comments and helps.
This work is supported by the National Basic Research Program (``973'' Program) of China (grant Nos.
2014CB845800 and 2013CB834900), the program A for Outstanding PhD  candidate of Nanjing University,
and the National Natural Science Foundation of China
(grant Nos. 11033002, 11422325, 11373022 and 11322328). X.F.W was also partially supported by the One-
Hundred-Talent Program, the Youth Innovation Promotion Association, and the Strategic
Priority Research Program ``The Emergence of Cosmological Structure'' (grant No. XDB09000000)
of the Chinese Academy of Sciences, and the Natural Science Foundation of
Jiangsu Province (No. BK2012890). F.Y.W was also partially supported by the Excellent Youth Foundation
of Jiangsu Province (BK20140016).



\clearpage
\begin{figure}[t]
 \centering
 \includegraphics[angle=0,scale=0.35,width=0.6\textwidth,height=0.4\textheight]{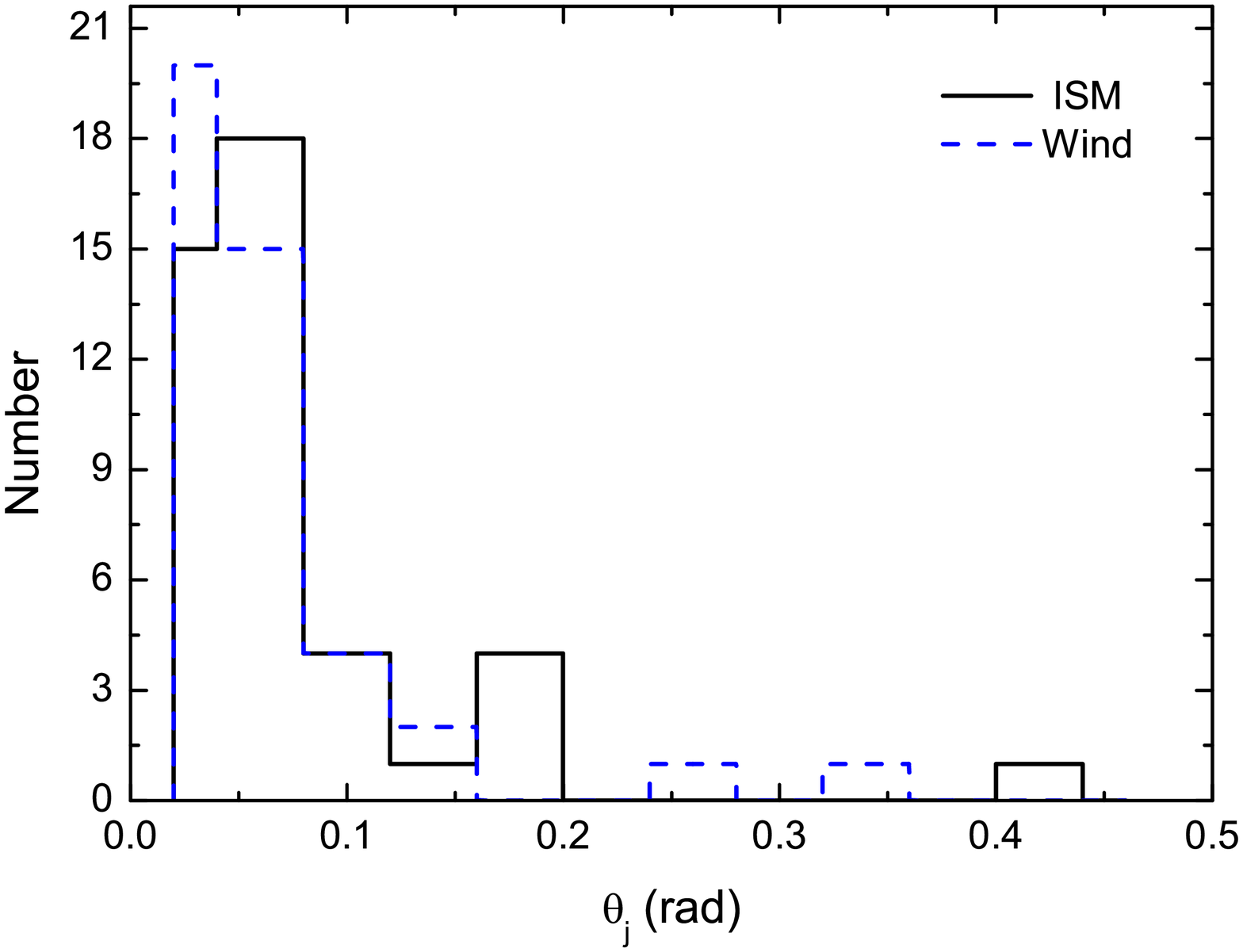}
 \caption{The distribution of half-opening angles of GRB jets. The solid and dash lines are corresponding to the ISM and wind cases, respectively.}
\end{figure}

\begin{figure}[t]
 \centering
 \includegraphics[angle=0,scale=0.35,width=0.6\textwidth,height=0.4\textheight]{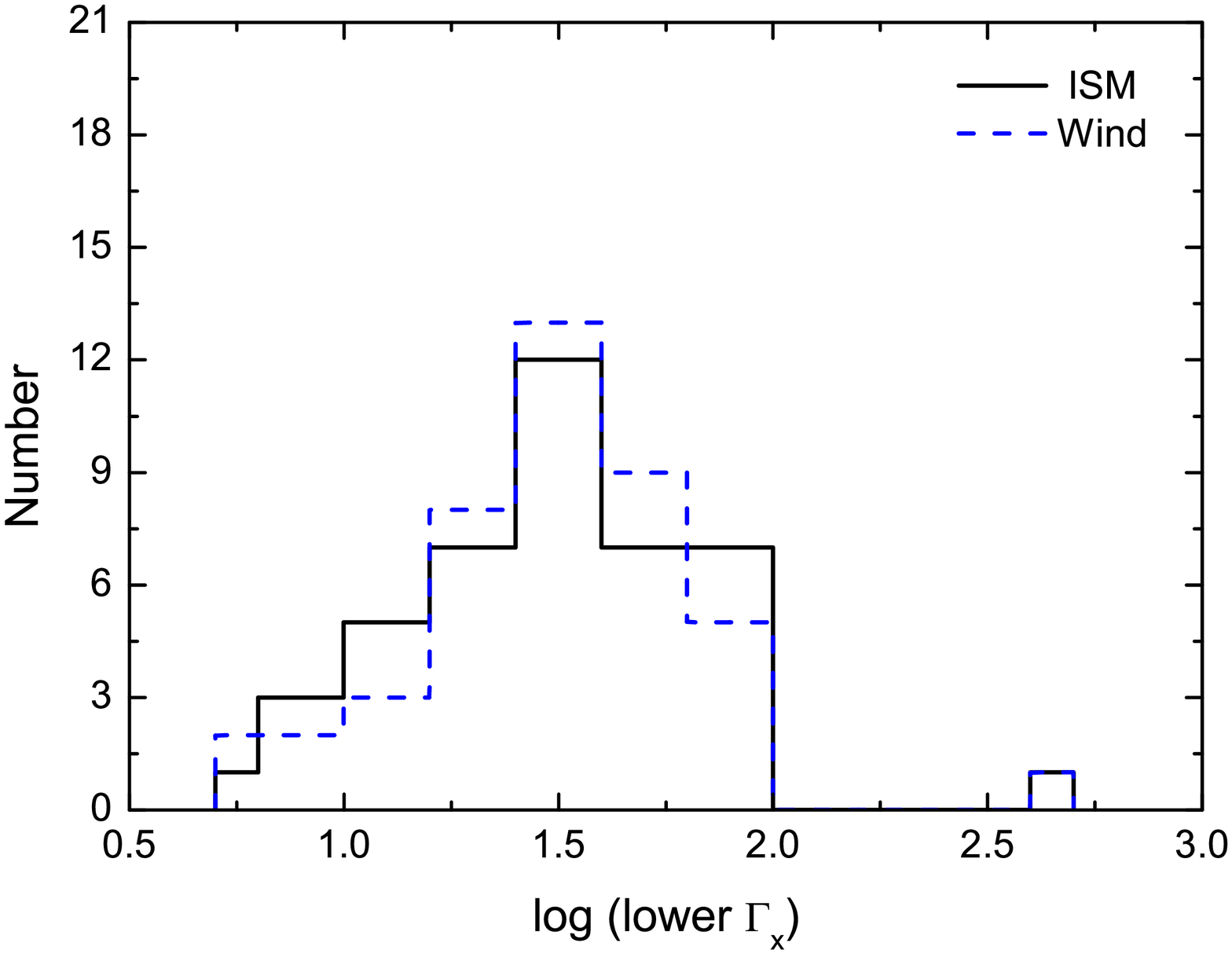}
 \caption{The distribution of the lower limits on the Lorentz
  factor of X-ray flares for two types of circumburst environment. The lower limits on $\Gamma_{x}$ range from tens to a few hundreds for the ISM (solid) and wind (dashed) cases. }
\end{figure}

\begin{figure}[t]
 \centering
 \includegraphics[angle=0,scale=0.35,width=0.6\textwidth,height=0.4\textheight]{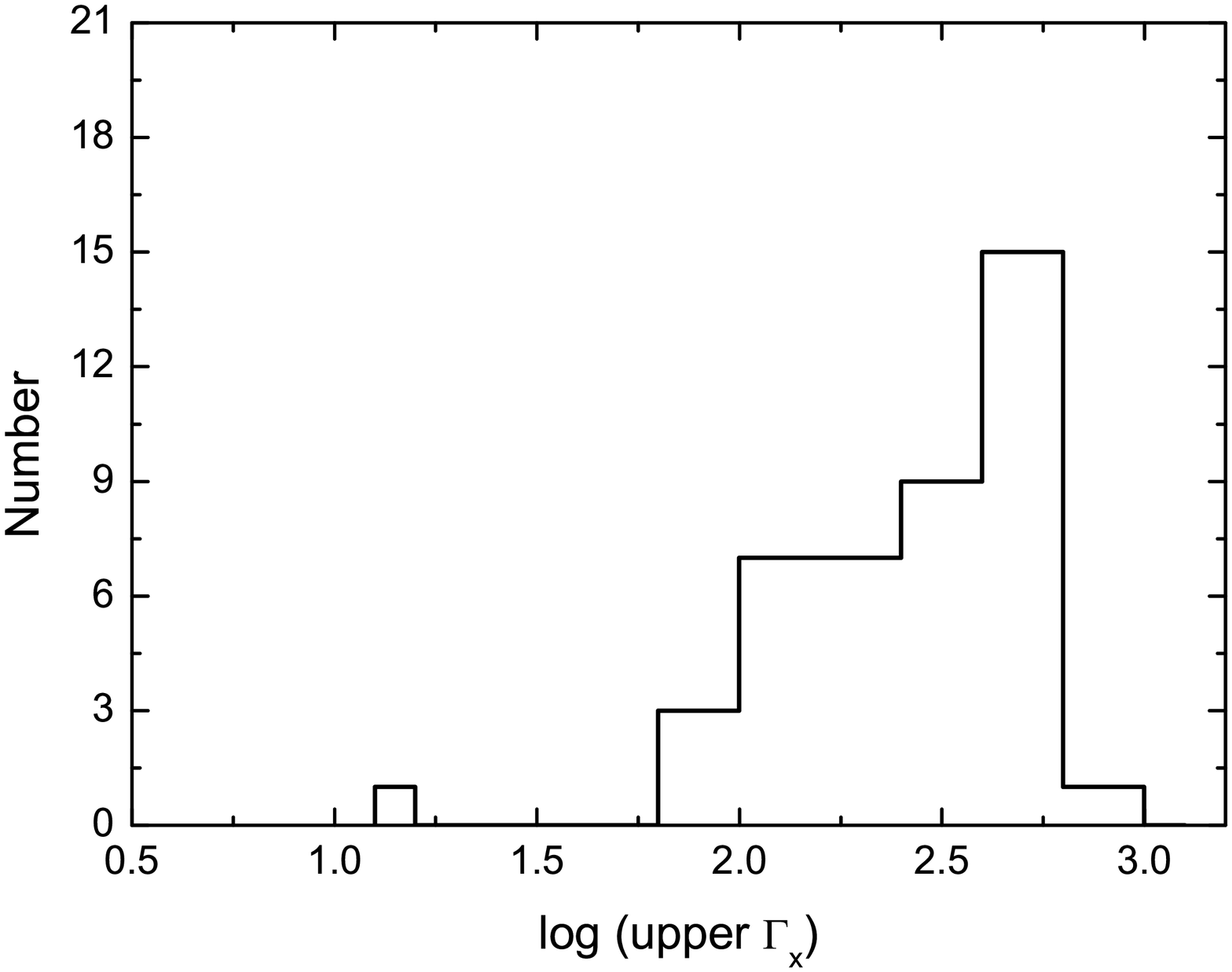}
 \caption{The distribution of the upper limits on the X-ray flare Lorentz factor.   }
\end{figure}

\begin{figure}[t]
 \centering
 \includegraphics[angle=0,scale=0.35,width=0.6\textwidth,height=0.4\textheight]{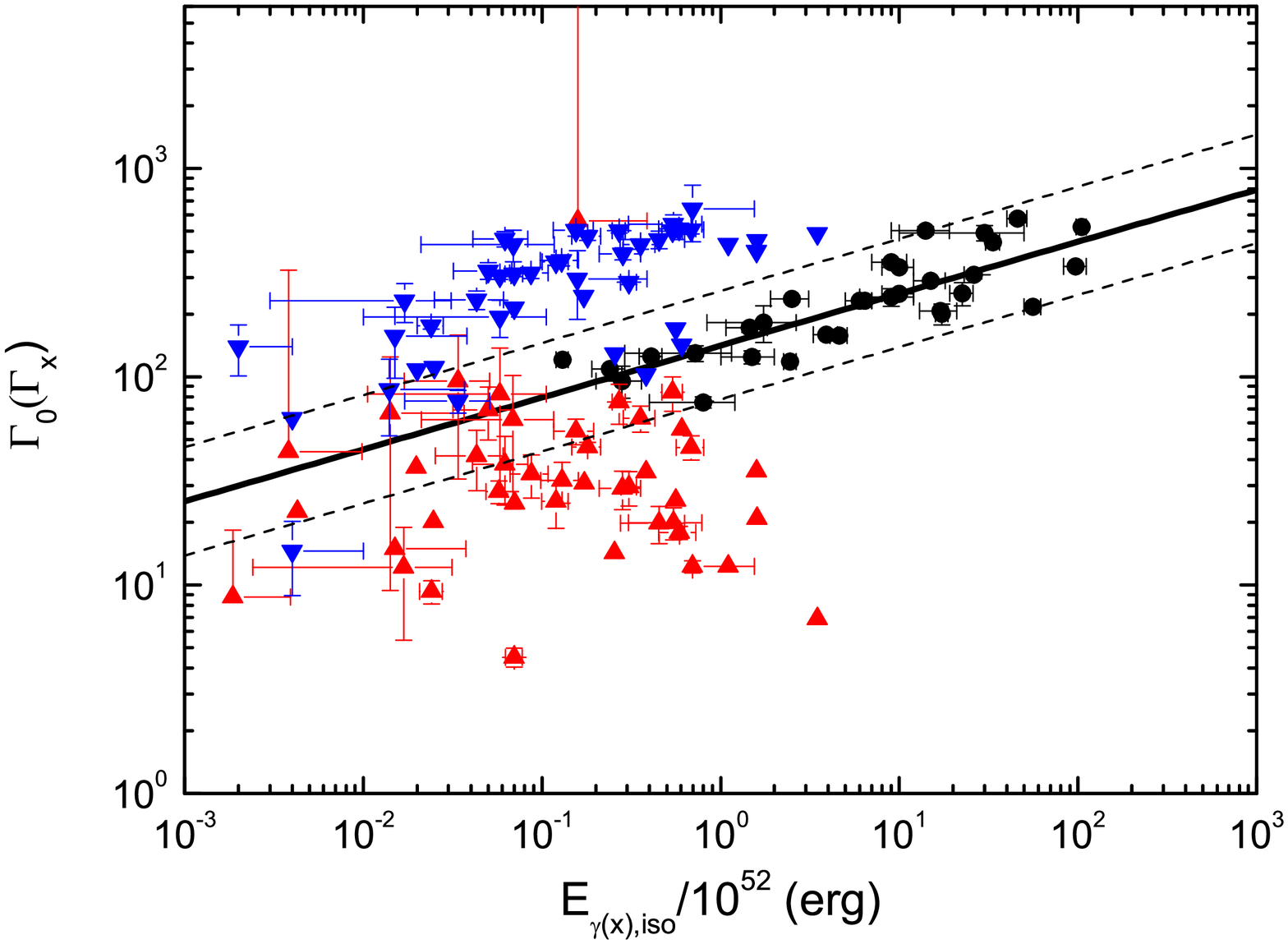}
 \caption{Lorentz  factor  and  isotropic  radiation energy of  X-ray flares
 (rest frame 0.3-10 keV, triangles) and prompt GRBs (solid dots) for the ISM case. The best fit power-law index of the correlation for prompt GRBs is 0.25. }
\end{figure}

\begin{figure}[t]
 \centering
 \includegraphics[angle=0,scale=0.35,width=0.6\textwidth,height=0.4\textheight]{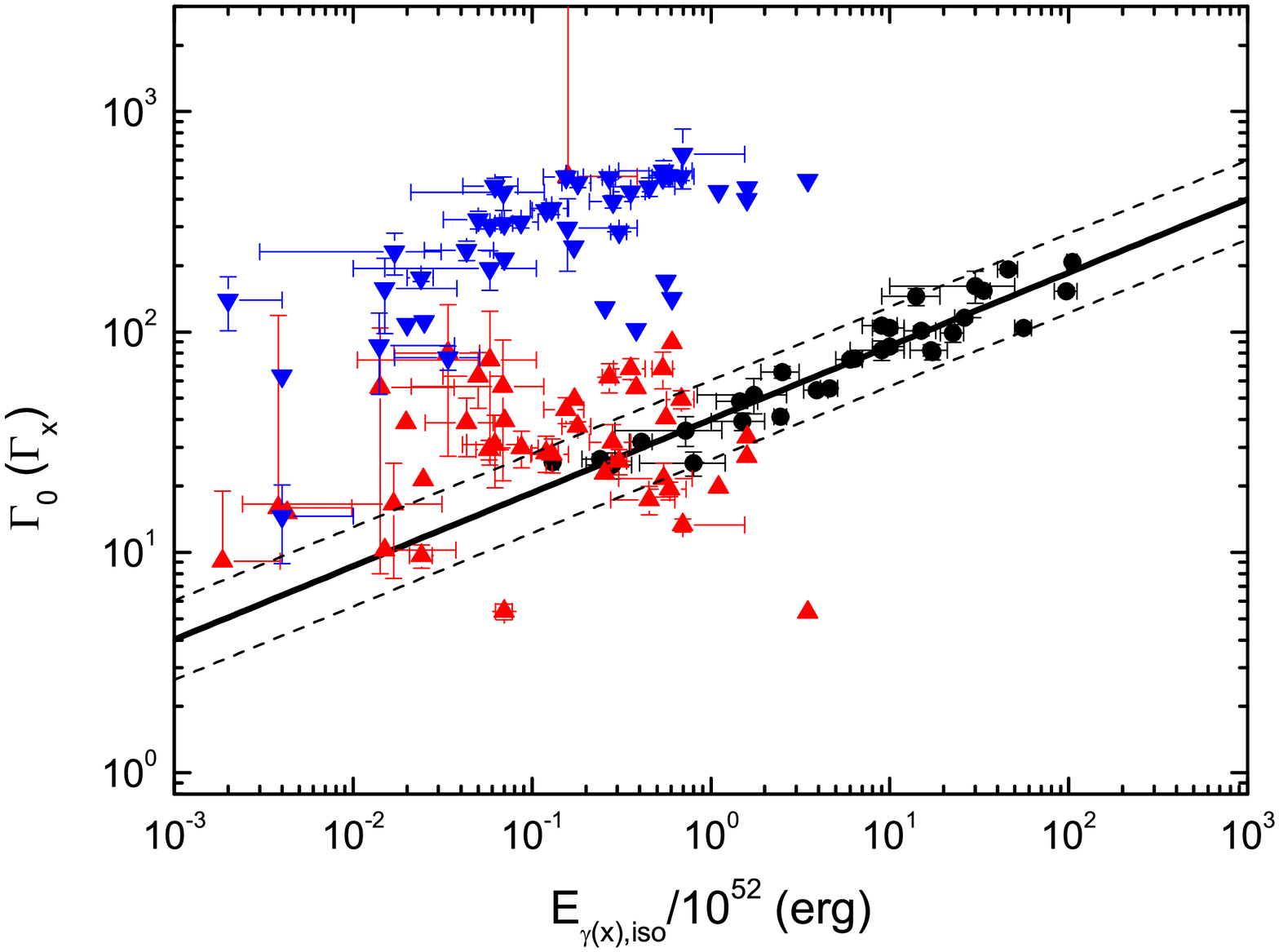}
 \caption{Lorentz  factor  and  isotropic  radiation energy of  X-ray flares (rest frame 0.3-10 keV, triangles) and prompt GRBs (solid dots)
for the wind case. The best fit power-law index of the correlation for prompt GRBs is 0.33.}
\end{figure}

\begin{figure}[t]
 \centering
 \includegraphics[angle=0,scale=0.35,width=1.0\textwidth,height=0.33\textheight]{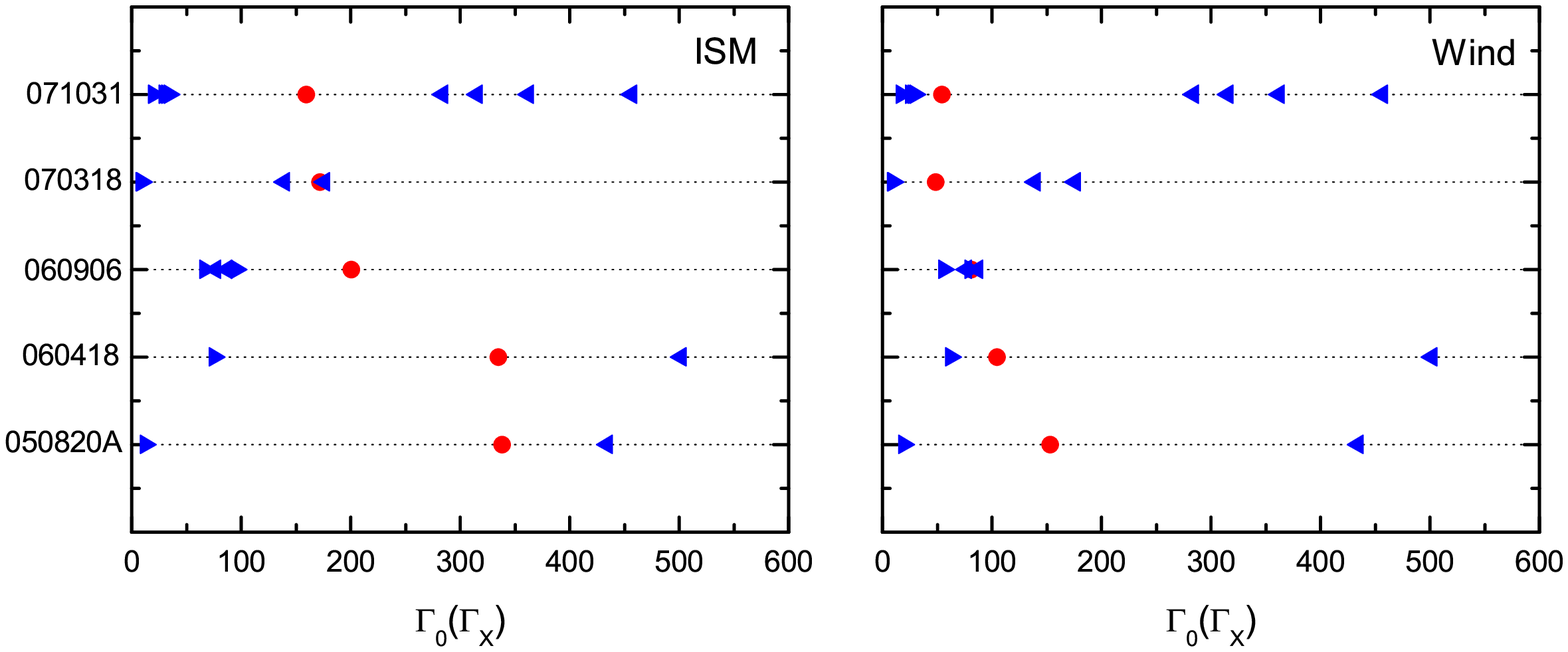}
 \caption{Prompt (solid dots) and X-ray flare (triangles) Lorentz factors of 5 GRBs.}
\end{figure}



\clearpage
\begin{deluxetable}{lcccccccccccccc}
\rotate
\tablewidth{650pt}
\tabletypesize{\tiny}
\tablecaption{Properties of GRB X-ray Flares}
\tablenum{1}

\tablehead{
 \colhead{GRB}&\colhead{z}&\colhead{$T_{rise}$}&\colhead{$T_{decay}$}&\colhead{$S_x$}&\colhead{$T_{90,x}$}&
\colhead{$E_{x,iso,50}$}&\colhead{$\theta_{j}^{ISM}$}&\colhead{$\theta_{j}^{Wind}$}&\colhead{Lower $\Gamma_{x}^{ISM}$}
&\colhead{Lower $\Gamma_{x}^{Wind}$}&\colhead{Upper $\Gamma_{x}$}
\\  \colhead{}&\colhead{}&\colhead{(s)}&\colhead{(s)}&\colhead{($10^{-7}$ erg cm$^{-2}$)}&\colhead{(s)}&\colhead{($10^{50} $erg)}&\colhead{(rad)}&\colhead{(rad)}&\colhead{}&\colhead{}&\colhead{} }

\startdata
050416A     &   0.650   &   2.2E5   $\pm$   2.8E6   &   2.8E5   $\pm$   5.8E5   &   0.34    $\pm$   0.53    &   5.0E5   &   0.4 $\pm$   0.6 &   0.026    $\pm$  0.005   &   0.071   $\pm$   0.009   &   43.7    $\pm$   281.9   &   15.9    $\pm$   102.7   &   14.6    $\pm$   5.7 &\\
050802      &   1.710   &   123.0   $\pm$   0       &   22.0    $\pm$   0       &   0.20    $\pm$   0.30    &   145.0   &   1.5 $\pm$   2.3 &   0.028    $\pm$  0.002   &   0.041   $\pm$   1E-3    &   14.9    $\pm$   0   &   10.3    $\pm$   0   &   157.1   $\pm$   58.9    &\\
050814 (1)  &   5.300   &   217.0   $\pm$   0   &   624.0   $\pm$   0   &   0.04    $\pm$   0   &   841.0   &   2.0 $\pm$   0   &   0.046   $\pm$    0.003          &   0.044   $\pm$   0.002   &   36.6    $\pm$   0   &   38.7    $\pm$   0   &   108.5   $\pm$   0   &\\
050814 (2)  &   5.300   &   505.0   $\pm$   0   &   439.0   $\pm$   0   &   0.05    $\pm$   0   &   944.0   &   2.5 $\pm$   0   &   0.046   $\pm$    0.003          &   0.044   $\pm$   0.002   &   20.1    $\pm$   0   &   21.3    $\pm$   0   &   111.4   $\pm$   0   &\\
050820A     &   2.620   &   34.0    $\pm$   0   &   148.0   $\pm$   0   &   6.89    $\pm$   0   &   182.0   &   110.5   $\pm$   0   &   0.169   $\pm$    0.007  &   0.106   $\pm$   0.003   &   12.3    $\pm$   0   &   19.7    $\pm$   0   &   434.5   $\pm$   0   &\\
050904 (1)  &   6.290   &   120.0   $\pm$   0   &   107.0   $\pm$   0   &   2.51    $\pm$   0   &   227.0   &   160.3   $\pm$   0   &   0.045   $\pm$    0.007  &   0.028   $\pm$   0.003   &   20.8    $\pm$   0   &   33.2    $\pm$   0   &   451.3   $\pm$   0   &\\
050904 (2)  &   6.290   &   96.0    $\pm$   0   &   188.0   $\pm$   0   &   0.27    $\pm$   0   &   284.0   &   17.2    $\pm$   0   &   0.045   $\pm$    0.007  &   0.028   $\pm$   0.003   &   30.9    $\pm$   0   &   49.2    $\pm$   0   &   244.4   $\pm$   0   &\\
050904 (3)  &   6.290   &   86.0    $\pm$   0   &   108.0   $\pm$   0   &   0.11    $\pm$   0   &   194.0   &   7.0 $\pm$   0   &   0.045   $\pm$    0.007  &   0.028   $\pm$   0.003   &   24.7    $\pm$   0   &   39.4    $\pm$   0   &   214.7   $\pm$   0   &\\
050904 (4)  &   6.290   &   1680.0  $\pm$   0   &   2236.0  $\pm$   0   &   0.88    $\pm$   0   &   3916.0  &   56.2    $\pm$   0   &   0.045   $\pm$    0.007  &   0.028   $\pm$   0.003   &   25.4    $\pm$   0   &   40.6    $\pm$   0   &   170.4   $\pm$   0   &\\
050904 (5)  &   6.290   &   1176.0  $\pm$   0   &   7537.0  $\pm$   0   &   0.95    $\pm$   0   &   8713.0  &   60.7    $\pm$   0   &   0.045   $\pm$    0.007  &   0.028   $\pm$   0.003   &   55.9    $\pm$   0   &   89.1    $\pm$   0   &   142.2   $\pm$   0   &\\
050904 (6)  &   6.290   &   5773.0  $\pm$   0   &   14457.0 $\pm$   0   &   0.60    $\pm$   0   &   20230.0 &   38.3    $\pm$   0   &   0.045   $\pm$    0.007  &   0.028   $\pm$   0.003   &   34.9    $\pm$   0   &   55.7    $\pm$   0   &   102.7   $\pm$   0   &\\
050904 (7)  &   6.290   &   3774.0  $\pm$   0   &   1586.0  $\pm$   0   &   0.40    $\pm$   0   &   5360.0  &   25.6    $\pm$   0   &   0.045   $\pm$    0.007  &   0.028   $\pm$   0.003   &   14.3    $\pm$   0   &   22.8    $\pm$   0   &   129.3   $\pm$   0   &\\
051016B &   0.940   &   109.0   $\pm$   0   &   1457.0  $\pm$   0   &   0.18    $\pm$   0   &   1566.0  &   0.4 $\pm$   0   &   0.162   $\pm$   0.039    &  0.242   $\pm$   0.039   &   22.6    $\pm$   0   &   15.1    $\pm$   0   &   63.3    $\pm$   0   &\\
060115  &   3.530   &   19.1    $\pm$   10.2    &   64.3    $\pm$   9.3 &   0.17    $\pm$   0.07    &   83.4    &   4.3 $\pm$   1.8 &   0.044   $\pm$    0.007  &   0.047   $\pm$   0.005   &   41.8    $\pm$   13.4    &   38.7    $\pm$   11.5    &   234.5   $\pm$   24.1    &\\
060124 (1)  &   2.300   &   291.0   $\pm$   0   &   70.0    $\pm$   0   &   27.13   $\pm$   0   &   361.0   &   347.5   $\pm$   0   &   0.071   $\pm$    0.005  &   0.092   $\pm$   0.005   &   6.9 $\pm$   0   &   5.3 $\pm$   0   &   487.6   $\pm$   0   &\\
060124 (2)  &   2.300   &   50.0    $\pm$   0   &   313.0   $\pm$   0   &   12.40   $\pm$   0   &   363.0   &   158.8   $\pm$   0   &   0.071   $\pm$    0.005  &   0.092   $\pm$   0.005   &   35.2    $\pm$   0   &   27.2    $\pm$   0   &   400.3   $\pm$   0   &\\
060210 (1)  &   3.910   &   23.1    $\pm$   1.5 &   37.6    $\pm$   2.1 &   2.30    $\pm$   0.40    &   60.7    &   68.4    $\pm$   11.9    &   0.028    $\pm$  0.003   &   0.026   $\pm$   0.002   &   45.9    $\pm$   6.1 &   49.4    $\pm$   4.6 &   507.1   $\pm$   22.0    &\\
060210 (2)  &   3.910   &   14.7    $\pm$   1.8 &   45.5    $\pm$   3.2 &   1.20    $\pm$   0.24    &   60.2    &   35.7    $\pm$   7.1 &   0.028    $\pm$  0.003   &   0.026   $\pm$   0.002   &   63.3    $\pm$   9.1 &   68.1    $\pm$   7.4 &   431.9   $\pm$   21.6    &\\
060418  &   1.490   &   5.6 $\pm$   0.5 &   19.5    $\pm$   0.5 &   4.80    $\pm$   0.40    &   25.0    &   27.0    $\pm$   2.3 &   0.025   $\pm$    0.005  &   0.030   $\pm$   0.004   &   75.6    $\pm$   16.6    &   62.2    $\pm$   9.3 &   501.9   $\pm$   10.5    &\\
060526 (1)  &   3.220   &   11.8    $\pm$   0.6 &   12.7    $\pm$   1.6 &   3.20    $\pm$   3.90    &   24.5    &   69.6    $\pm$   84.9    &   0.0850    $\pm$  7.891E-4    &   0.0780   $\pm$   4.86E-4 &   12.3    $\pm$   0.8 &   13.3    $\pm$   0.9 &   639.1   $\pm$   194.7   &\\
060526 (2)  &   3.220   &   9.8 $\pm$   1.6 &   27.9    $\pm$   8.6 &   2.50    $\pm$   1.10    &   37.8    &   54.4    $\pm$   23.9    &   0.0850    $\pm$  7.891E-4    &   0.0780   $\pm$   4.86E-4 &   20.0    $\pm$   3.5 &   21.6    $\pm$   3.8 &   539.1   $\pm$   59.3    &\\
060526 (3)  &   3.220   &   15.2    $\pm$   1.2 &   34.6    $\pm$   4.3 &   2.70    $\pm$   0.63    &   49.9    &   58.7    $\pm$   13.7    &   0.0850    $\pm$  7.891E-4    &   0.0780   $\pm$   4.86E-4 &   17.9    $\pm$   1.3 &   19.3    $\pm$   1.4 &   512.7   $\pm$   29.9    &\\
060526 (4)  &   3.220   &   10.2    $\pm$   4.2 &   61.7    $\pm$   3.7 &   1.30    $\pm$   0.34    &   71.9    &   28.3    $\pm$   7.4 &   0.0850    $\pm$  7.891E-4    &   0.0780   $\pm$   4.86E-4 &   29.1    $\pm$   6.1 &   31.5    $\pm$   6.6 &   389.8   $\pm$   25.5    &\\
060707  &   3.425   &   8.1 $\pm$   8.0 &   26.5    $\pm$   10.0    &   0.07    $\pm$   0.06    &   34.6    &   1.7 $\pm$   1.4 &   0.149   $\pm$    0.024  &   0.110   $\pm$   0.012   &   12.2    $\pm$   6.7 &   16.5    $\pm$   8.9 &   231.3   $\pm$   49.6    &\\
060714 (1)  &   2.711   &   8.1 $\pm$   3.0 &   41.6    $\pm$   2.0 &   3.30    $\pm$   0.43    &   49.6    &   53.9    $\pm$   7.0 &   0.0270   $\pm$    8.414E-4   &   0.0330   $\pm$   6.938E-4    &   84.2    $\pm$   15.9    &   68.0    $\pm$   12.8    &   502.7   $\pm$   16.4    &\\
060714 (2)  &   2.711   &   4.0 $\pm$   2.1 &   4.2 $\pm$   2.1 &   0.38    $\pm$   0.13    &   8.2 &   6.2 $\pm$   2.1 &   0.0270   $\pm$   8.414E-4     &  0.0330   $\pm$   6.938E-4    &   38.1    $\pm$   13.9    &   30.8    $\pm$   11.2    &   459.2   $\pm$   39.3    &\\
060714 (3)  &   2.711   &   4.4 $\pm$   1.1 &   9.6 $\pm$   1.0 &   0.95    $\pm$   0.24    &   14.0    &   15.5    $\pm$   3.9 &   0.0270   $\pm$    8.414E-4   &   0.0330   $\pm$   6.938E-4    &   54.9    $\pm$   7.6 &   44.4    $\pm$   6.1 &   505.2   $\pm$   31.9    &\\
060714 (4)  &   2.711   &   8.3 $\pm$   0.6 &   12.8    $\pm$   0.5 &   1.10    $\pm$   0.20    &   21.1    &   18.0    $\pm$   3.3 &   0.0270   $\pm$    8.414E-4   &   0.0330   $\pm$   6.938E-4    &   46.1    $\pm$   2.4 &   37.3    $\pm$   1.7 &   473.0   $\pm$   21.5    &\\
060729  &   0.540   &   9.6 $\pm$   1.0 &   34.1    $\pm$   2.0 &   9.30    $\pm$   1.00    &   43.7    &   7.0 $\pm$   0.8 &   0.418   $\pm$   0.037    &  0.350   $\pm$   0.020   &   4.5 $\pm$   0.5 &   5.4 $\pm$   0.5 &   311.3   $\pm$   8.4 &\\
060814  &   0.840   &   9.3 $\pm$   1.3 &   29.7    $\pm$   2.3 &   3.10    $\pm$   0.48    &   39.0    &   5.8 $\pm$   0.9 &   0.064   $\pm$   0.006    &  0.061   $\pm$   0.004   &   28.1    $\pm$   3.5 &   29.2    $\pm$   3.0 &   305.2   $\pm$   11.8    &\\
060906 (1)  &   3.690   &   349.0   $\pm$   397.0   &   1110.0  $\pm$   1420.0  &   0.05    $\pm$   0.08    &   1460.0  &   1.4 $\pm$   2.3 &   0.027    $\pm$  0.002   &   0.032   $\pm$   1E-3    &   66.9    $\pm$   57.4    &   56.0    $\pm$   48.0    &   86.8    $\pm$   34.7    &\\
060906 (2)  &   3.690   &   774.0   $\pm$   987.0   &   5030.0  $\pm$   1640.0  &   0.12    $\pm$   0.06    &   5810.0  &   3.4 $\pm$   1.7 &   0.027    $\pm$  0.002   &   0.032   $\pm$   1E-3    &   95.6    $\pm$   63.3    &   80.1    $\pm$   52.9    &   76.5    $\pm$   9.6 &\\
070306  &   1.500   &   8.7 $\pm$   1.4 &   34.7    $\pm$   3.3 &   2.10    $\pm$   0.35    &   43.4    &   12.0    $\pm$   2.0 &   0.079   $\pm$    0.019  &   0.071   $\pm$   0.011   &   25.3    $\pm$   6.6 &   28.3    $\pm$   5.3 &   356.8   $\pm$   14.9    &\\
070318 (1)  &   0.840   &   9.5 $\pm$   16.2    &   19.4    $\pm$   26.1    &   0.10    $\pm$   0.11    &   28.9    &   0.2 $\pm$   0.2 &   0.163    $\pm$  0.011   &   0.157   $\pm$   0.007   &   8.8 $\pm$   9.6 &   9.1 $\pm$   9.9 &   139.4   $\pm$   38.3    &\\
070318 (2)  &   0.840   &   44.9    $\pm$   8.3 &   103.0   $\pm$   12.6    &   1.30    $\pm$   0.19    &   147.9   &   2.4 $\pm$   0.4 &   0.163    $\pm$  0.011   &   0.157   $\pm$   0.007   &   9.3 $\pm$   1.2 &   9.6 $\pm$   1.1 &   176.0   $\pm$   6.4 &\\
070721B (1) &   3.626   &   0.9 $\pm$   145.7   &   120.8   $\pm$   149.1   &   0.60    $\pm$   0.87    &   121.7   &   15.8    $\pm$   23.0    &    0.0210  $\pm$   7.033E-4    &   0.0230   $\pm$   5.182E-4    &   561.6   $\pm$   45459.9 &   508.2   $\pm$   41133.9 &   295.7   $\pm$   107.2   & \\
070721B (2) &   3.626   &   4.4 $\pm$   4.4 &   7.3 $\pm$   5.5 &   0.26    $\pm$   0.18    &   11.7    &   6.9 $\pm$   4.8 &   0.0210   $\pm$    7.033E-4   &   0.0230   $\pm$   5.182E-4    &   62.4    $\pm$   39.1    &   56.5    $\pm$   35.4    &   430.8   $\pm$   74.6    &\\
070721B (3) &   3.626   &   8.9 $\pm$   4.0 &   18.3    $\pm$   6.3 &   0.19    $\pm$   0.07    &   27.2    &   5.0 $\pm$   1.8 &   0.0210   $\pm$    7.033E-4   &   0.0230   $\pm$   5.182E-4    &   69.5    $\pm$   19.8    &   62.9    $\pm$   17.9    &   322.6   $\pm$   29.7    &\\
070721B (4) &   3.626   &   62.2    $\pm$   64.3    &   179.8   $\pm$   151.2   &   0.22    $\pm$   0.18    &   242.0   &   5.8 $\pm$   4.8 &   0.0210    $\pm$  7.033E-4    &   0.0230   $\pm$   5.182E-4    &   82.4    $\pm$   55.0    &   74.6    $\pm$   49.7    &   193.7   $\pm$   39.6    &\\
071031 (1)  &   2.690   &   25.0    $\pm$   3.5 &   36.2    $\pm$   2.6 &   2.80    $\pm$   1.10    &   61.2    &   45.2    $\pm$   17.8    &   0.061    $\pm$  0.011   &   0.069   $\pm$   0.009   &   19.8    $\pm$   4.0 &   17.3    $\pm$   2.5 &   456.4   $\pm$   44.8    &\\
071031 (2)  &   2.690   &   9.3 $\pm$   1.9 &   34.7    $\pm$   5.0 &   0.80    $\pm$   0.19    &   44.0    &   12.9    $\pm$   3.1 &   0.061   $\pm$    0.011  &   0.069   $\pm$   0.009   &   31.8    $\pm$   7.1 &   27.8    $\pm$   4.9 &   362.4   $\pm$   21.5    &\\
071031 (3)  &   2.690   &   9.8 $\pm$   2.4 &   42.1    $\pm$   6.3 &   0.54    $\pm$   0.13    &   51.9    &   8.7 $\pm$   2.1 &   0.061   $\pm$    0.011  &   0.069   $\pm$   0.009   &   34.1    $\pm$   8.0 &   29.8    $\pm$   5.7 &   315.2   $\pm$   19.0    &\\
071031 (4)  &   2.690   &   65.5    $\pm$   4.4 &   210.6   $\pm$   9.0 &   1.90    $\pm$   0.19    &   276.1   &   30.7    $\pm$   3.1 &   0.061    $\pm$  0.011   &   0.069   $\pm$   0.009   &   29.5    $\pm$   5.6 &   25.8    $\pm$   3.3 &   284.3   $\pm$   7.1 &\\
\enddata

\end{deluxetable}


\end{document}